\documentstyle[aps,preprint,epsf]{revtex}
\newcommand{\beq}{\begin{equation}}
\newcommand{\eeq}{\end{equation}}
\begin{document}
\draft
\tightenlines

\title{ Behavior of Fermi Systems Approaching
Fermion Condensation Quantum
Phase Transition from Disordered Phase}

\author{ V.R. Shaginyan$^{a,b}$\footnote{E--mail:
vrshag@thd.pnpi.spb.ru}}
\address{ $^a$Petersburg Nuclear Physics
Institute, Russian Academy of Sciences, Gatchina,
188300, Russia;\\
$^b\,$CTSPS, Clark Atlanta University,
Atlanta, Georgia 30314, USA}
\maketitle

\begin{abstract}

The behavior of Fermi systems which approach
the fermion condensation quantum phase transition (FCQPT)
from the disordered phase is considered.
We show that the quasiparticle effective mass $M^*$
diverges as $M^*\propto 1/|x-x_{FC}|$ where $x$
is the system density and $x_{FC}$ is the
critical point at which FCQPT occurs. Such a behavior is
of general form and takes place in both three dimensional (3D)
systems and two dimensional (2D) ones.
Since the effective mass $M^*$ is finite, the
system exhibits the Landau Fermi liquid behavior.
At $|x-x_{FC}|/x_{FC}\ll 1$, the behavior can be viewed as a
highly correlated one, because the effective mass is large and
strongly depends on the density.
In case of electronic systems the Wiedemann-Franz law is held
and Kadowaki-Woods ratio is preserved.
Beyond the region $|x-x_{FC}|/x_{FC}\ll 1$, the
effective mass is approximately constant
and the system becomes  conventional Landau Fermi liquid.
\end{abstract}\bigskip


It is widely believed that unusual properties of the strongly
correlated liquids observed in the high-temperature
superconductors, heavy-fermion metals, 2D $^3$He and etc., are
determined by quantum phase transitions.
Any quantum phase transition occurs
at temperature $T=0$ and is driven by a control parameter other
then temperature, for instance, by pressure, by magnetic field,
or by the density $x$.
A quantum phase transition occurs at the
quantum critical point. As any phase transition, the
quantum phase transition is related to the order parameter which
induces a broken symmetry.
Therefore, direct experimental studies of relevant quantum phase
transitions are of crucial importance for
understanding the physics of the high-temperature
superconductivity and strongly correlated systems.

In case of the high-temperature superconductors, these experiments
are difficult to carry out, because at low temperatures
all the corresponding area is occupied by the
superconductivity. On the other hand, experimental data on
the behavior of different Fermi liquids, when systems are
approaching  the critical point from the disordered phase,
can help to illuminate both the nature of this point
and the nature of control parameter by which this phase transition
is driven. Experimental
facts on high-density  2D $^3$He
\cite{mor,cas} show that the effective mass diverges when the density
at which 2D $^3$He liquid begins to solidify is approached
\cite{cas}. Then, the sharp increase of the effective mass
in a metallic 2D electron system is observed
when the
density tends to the critical density of the metal-insulator
transition point.  This transition occurs at sufficiently
low densities \cite{skdk}.
Note, that
there is no ferromagnetic instability in both Fermi systems and the
relevant Landau amplitude $F^a_0>-1$ \cite{cas,skdk}, in accordance
with the almost localized fermion model \cite{pfw}.

Recent measurements for
non-superconducting La$_{1.7}$Sr$_{0.3}$CuO$_4$
have shown that the resistivity $\rho$ exhibits $T^2$ behavior,
$\rho=\rho_0+\Delta\rho$ with $\Delta\rho=AT^2$, that the
Wiedemann-Franz (WF) law is verified
to hold perfectly, and that the Kadowaki-Woods ratio,
$A/\gamma_0^2$ \cite{kadw},
is enhanced compared with heavy-fermion metals
\cite{nakam}. Here $\gamma_0$ is the linear specific heat
coefficient, $C=\gamma_0T$. These data demonstrate the behavior of the
Fermi liquid located above the critical point, or on the side of the
disordered phase.

In this Letter, we study the behavior of Fermi systems which approach
the fermion condensation quantum phase transition (FCQPT) \cite{ms}
from disordered phase and show that the outlined experimental
data can be explained within the framework of our approach.
We analyze the appearance of FCQPT
in different 2D and 3D Fermi liquids and
show that at $T\to 0$ FCQPT manifests itself in the divergence of the
quasiparticle effective mass $M^*$ as the density $x$ of a system
approaches the critical point $x_{FC}$ at which FCQPT takes place, so
that $M^*\propto 1/|x-x_{FC}|$.
Since the effective mass $M^*$ is finite, the
system exhibits the Landau Fermi liquid (LFL) behavior
at low temperatures. At sufficiently high temperatures,
the system possesses the non Fermi liquid (NFL) behavior.
At $|x-x_{FC}|/x_{FC}\ll 1$,
this behavior can be viewed as
a highly correlated one, because the effective mass
strongly depends on the density and is large.
We show that in the case of electronic systems the WF law is held,
and Kadowaki-Woods ratio is preserved.
Beyond the region $|x-x_{FC}|/x_{FC}\ll 1$,
the effective mass is approximately constant
and the system becomes conventional Landau Fermi liquid.

A new state of Fermi liquid with the fermion condensate (FC)
\cite{ks,vol} which
takes place beyond the critical point $x_{FC}$ is defined by the
equation \cite{ks}
\beq \frac{E[n(p)]}{\delta n({\bf p})}-\mu=
\varepsilon({\bf p})-\mu=0,\,\,\: {\mathrm {if}}\,
0<n({\bf p})<1;
\,\: p_i\leq p\leq p_f\in L_{FC}.\eeq
Here $E[n(p)]$ is the Landau functional of the ground state energy,
$n({\bf p})$ is the quasiparticle distribution function,
$\varepsilon({\bf p})$ is the single-particle
energy of the quasiparticles, and $\mu$ is
the chemical potential \cite{lan}. At $T=0$, Eq. (1)
defines a new state of Fermi liquid with the fermion condensate (FC)
for which the modulus of the superconducting order
parameter $|\kappa({\bf p})|$ has
finite values in $L_{FC}$ range of momenta $p_i\leq p\leq p_f$
occupied by FC. At the same time the superconducting gap can be
infinitely small, $\Delta_1\to 0$ in $L_{FC}$ provided
the corresponding pairing interaction is also small
\cite{ms,ks}. Such a state can be considered as
superconducting, with infinitely small value of $\Delta_1$ so that
the entropy of this state is equal to zero. This
state, created by the quantum phase transition, disappears at
$T>0$. FCQPT can be considered as a ``pure'' quantum phase transition
because it cannot take place at finite temperatures.
Therefore, the corresponding quantum critical point does not represent
the end of a line of continuous phase transitions at $T=0$.
Nonetheless, FCQPT has a
strong impact on the system's properties up to
temperature $T_f$ above which FC effects become insignificant
\cite{ms,ks}. FCQPT does not violate any rotational symmetry or
translational symmetry, being characterized by the
order parameter
$\kappa({\bf p})=\sqrt{n({\bf p})(1-n({\bf p}))}$.
It follows from Eq. (1) that the
quasiparticle system splits on the two quasiparticle subsystems: the
first one in $L_{FC}$ range is occupied by the quasiparticles with
the effective mass $M^*_{FC}\propto 1/\Delta_1$, while the second by
quasiparticles with finite mass $M^*_L$ and momenta $p<p_i$.
Note, that the existence of such a state can be revealed
experimentally. Since the order parameter
$\kappa({\bf p})$ is suppressed
by a magnetic field $B$, when $B^2\sim \Delta_1^2$,
a weak magnetic field $B$ will destroy the state with FC converting
the strongly correlated Fermi liquid into the
normal Landau Fermi liquid.

Equation (1), possessing
solutions at some density $x=x_{FC}$,
determines the critical point of FCQPT.
It is also evident from Eq. (1)
that the effective mass diverges when $x\to x_{FC}$,
$M^*_L(x\to x_{FC})\to \infty$.
Let us assume that FC has just taken place,
$p_i\to p_f\to p_F$,
and the deviation $\delta n(p)$ is small.
Expanding functional $E[n(p)]$ in Taylor's series with
respect to $\delta n(p)$ and retaining the leading terms,
one obtains,
\beq \mu=\varepsilon({\bf p},\sigma) =
\varepsilon_0({\bf p},\sigma)+\sum_{\sigma_1}\int
F_L({\bf p},{\bf p}_1,\sigma,\sigma_1)
\delta n({\bf p_1},\sigma_1)
\frac{d{\bf p}_1}{(2\pi)^2};\,\,\:
p_i\leq p\leq p_f\in L_{FC}.\eeq
In Eq. (2)
$F_L({\bf p},{\bf p}_1,\sigma,\sigma_1)=
\delta^2 E/\delta n({\bf
p},\sigma)\delta n({\bf p}_1,\sigma_1)$
is the Landau interaction \cite{lan},
and $\sigma$ denotes the spin states.
Both the Landau interaction
and the single-particle energy
$\varepsilon_0(p)$ are calculated at
$n(p)=n_F(p)$. Here $n_F(p)=\theta(p_F-p)$, and
$\theta(p-p_F)$ is the Fermi-Dirac distribution at $T=0$.
Equation (2) possesses solutions when  the
Landau amplitude $F_L$ is positive and sufficiently large, so that
the integral on the right hand side of Eq. (2) defining the potential
energy is large and therefore the potential energy prevails over the
kinetic energy $\varepsilon_0({\bf p})$ \cite{ks}.
At temperatures $T\geq T_c$, the effective mass
$M^*_{FC}$ related to FC is given by \cite{ms,ams}, \beq
M^*_{FC}\simeq p_F\frac{p_f-p_i}{4T}.\eeq Multiplying both sides of
Eq. (3) by $p_f-p_i$ we obtain the energy scale $E_0$ separating the
slow dispersing low energy part related to the effective mass
$M^*_{FC}$, from the faster dispersing relatively high energy part
defined by the effective mass $M^*_{L}$ \cite{ms,ams},
\beq E_0\simeq 4T.\eeq It is clear from Eq.
(4) that the scale $E_0$ does not depend
on the range $p_f-p_i$.
It is natural to assume that we have returned back to the Landau
theory by integrating out high energy degrees of freedom and
introducing the quasiparticles. Sole difference between the Landau
Fermi liquid and Fermi liquid undergone FCQPT is that we have to
expand the number of relevant low energy degrees of freedom by
introducing new type of quasiparticles with the effective mass
$M^*_{FC}$ given by Eq. (3) and the energy scale $E_0$ given by Eq.
(4).  It is seen from Eqs. (1) and (2) that the FC quasiparticles
form a collective state, since their energies are defined by the
macroscopical number of quasiparticles within the region $p_i-p_f$.
The shape of the spectra is not affected by the Landau interaction,
which, generally speaking, depends on the system's properties,
including the collective states, impurities, etc. The only thing
defined by the interaction is the width of the region $p_i-p_f$,
provided the interaction is sufficiently strong to produce the FC
phase transition at all. The spectra related to FC are of universal
form and determined by $T$, as it follows from Eq. (3).  Thus, the
system's properties and dynamics are dominated by a strong collective
effect  originated from FCQPT and determined by the macroscopic
number of quasiparticles in the range $L_{FC}$.  Such a system can
viewed as a strongly correlated system and cannot be disturbed by the
scattering of individual quasiparticles, thermal excitations,
impurities, and etc, and has features of a quantum protectorate
\cite{ms,lpa,pwa}.

The appearance of FCQPT in Fermi liquids, when the effective
interaction becomes sufficiently large, has been predicted in
Ref.\cite{ksz}.  FCQPT precedes the formation of charge-density waves
or stripes, which take place at some value $x=x_{cdw}$ with
$x_{FC}>x_{cdw}$, while the Wigner solidification takes place even at
lower values of $x$ and leads to an insulator.  In the same way, the
effective mass inevitably diverges as soon as the density $x$ becomes
sufficiently large approaching the critical density at which 2D
$^3$He begins to solidify, as it was observed in
\cite{cas}.

Now we consider the divergence of the effective mass in 2D and 3D
Fermi liquids at $T=0$, when the density $x$ approaches FCQPT from
the side of  the normal LFL.
First, we calculate the divergence of $M^*$ as a function of the
difference $(x_{FC}-x)$ in case of 2D $^3$He.
For this purpose we use
the equation for $M^*$ obtained in Ref.\cite{ksz} where
the divergence of effective mass $M^*$ due to the onset of FC in
different Fermi liquids including $^3$He was predicted
\cite{ksz}
\beq\frac{1}{M^{*}}=\frac{1}{M}+\frac{1}{4\pi^{2}}
\int\limits_{-1}^{1}\int\limits_0^{g_0}
\frac{v(q(y))}{\left[1-R(q(y),\omega=0,g)
\chi_0(q(y),\omega=0)\right]^{2}}
\frac{ydydg}{\sqrt{1-y^{2}}}.
\eeq
Here we adopt the shorthand, $p_F\sqrt{2(1-y)}=q(y)$, with $q(y)$
is the transferred momentum, $M$ is the bare mass, $\omega$ is
the frequency, $v(q)$ is the bare interaction, and the integral is
taken over the coupling constant $g$ from zero to its real value
$g_0$. In Eq. (5), both $\chi_0(q,\omega)$ and $R(q,\omega)$,
being the linear response function of noninteracting Fermi liquid
and the effective interaction respectively, define the linear
response function of the system under consideration \beq
\chi(q,\omega,g)=\frac{\chi_0(q,\omega)}
{1-R(q,\omega,g)\chi_0(q,\omega)}.
\eeq
In the vicinity of charge density wave instability, occurring at the
density $x_{cdw}$, the singular part of the function $\chi^{-1}$ on
the disordered side is of the well-known form, see.  e.g.  \cite{vn}
\beq\chi^{-1}(q,\omega,g)\propto (x_{cdw}-x)+(q-q_c)^2+(g_0-g),\eeq
where $q_c\sim 2p_F$ is the wavenumber of the charge density wave.
Upon substituting Eq. (7) into Eq. (5) and performing the
integrations, the equation for the effective mass $M^*$ can be cast
into the following form \cite{shag}
\beq\frac{1}{M^*}=
\frac{1}{M}-\frac{C}{\sqrt{x_{cdw}-x}},\eeq with $C$ being some
positive constant. It is seen form Eq. (8) that $M^*$ diverges at
some point $x_{FC}$, which is referred to as the critical point,
as a function of the difference $(x_{FC}-x)$ \beq
M^*\sim M\frac{x_{FC}}{x_{FC}-x}.\eeq
It follows from the derivation of Eqs. (8) and (9) that the form of
these equations is independent from the bare interaction $v(q)$,
therefore both of these equations are also applicable to 2D electron
liquid or to another Fermi liquid. It is also seen from Eqs. (8) and
(9) that FCQPT precedes the formation of charge-density waves. As
consequence of this, the effective mass diverges at high densities in
case of 2D $^3$He, and it diverges at low density in case of 2D
electron systems, in accordance with experimental facts
\cite{cas,skdk}. Note, that in the both cases the difference
$(x_{FC}-x)$ has to be positive because $x_{FC}$ represents the
solution  of Eq. (8). Thus, considering electron systems we have to
replace $(x_{FC}-x)$ by $(x-x_{FC})$. In case of 3D system, the
effective mass is given by \cite{ksz}
\beq\frac{1}{M^{*}}=\frac{1}{M}+\frac{p_F}{4\pi^{2}}
\int\limits_{- 1}^{1}\int\limits_0^{g_0}
\frac{v(q(y))ydydg}{\left[1-R(q(y),\omega=0,g)
\chi_0(q(y),\omega=0)\right]^{2}}.
\eeq
A comparison of Eq. (10) and Eq. (5) shows that there is no
fundamental difference between these equations, and along
the same lines we again arrive to Eqs. (8) and (9). The only
difference between 2D electron systems and 3D ones is that FCQPT
occurs at densities which are well below those corresponding to 2D
systems. While in the bulk $^3$He, FCQPT cannot probably take place
being absorbed by the first order solidification.

Deriving Eq. (9), we assumed that the temperature $T=0$.
It is seen from Eq. (3) that the effective mass decreases
when the temperature increases. The same is true when the system
lies above the critical point.
Therefore, when
$T$ exceeds some temperature $T^*(x)$, Eq. (9) is no longer valid,
and $M^*$ depends on the temperature as well.
To estimate $T^*(x)$, we can compare the deviation
$\Delta x=|x-x_{FC}|$ with the deviation $\Delta x(T)$, generated
by $T$. The deviation $\Delta x$ can be
expressed in  terms of $M^*(x)$ using Eq. (9),
$\Delta x/x\sim M/M^*(x)$. On the other hand,
the temperature smoothing out the Fermi function $\theta(p_F-p)$
at $p_F$ induces the variation $p_F \Delta p/M^*(x)\sim T$. As a
result, we have $\Delta x(T)/x\sim M^*(x)T/p_F^2$. Comparing these
deviations, we find that at $T\geq T^*(x)$  the effective mass
depends noticeably on the temperature, and the equation for $T^*(x)$
becomes \beq T^*(x)\sim p_F^2\frac{M}{(M^*(x))^2}\sim
\varepsilon_F(x) \left(\frac{M}{M^*(x)}\right)^2.\eeq Here
$\varepsilon_F(x)$ is the Fermi energy of noninteracting electrons
with mass $M$.  From Eq. (11) it follows that $M^*$ is always finite
provided $T>0$. We can consider $T^*(x)$ as the energy scale
$e_0(x)\simeq T^*(x)$. This scale defines the area
$(\mu -e_0(x))$ in the single
particle spectrum where $M^*$ is approximately constant, being given
by $M^*=d\varepsilon(p)/dp$ \cite{lan}. According  to Eqs. (9) and
(11) it is easily verified that $e_0(x)$ can be written in the form
\beq e_0(x)\sim \varepsilon_F
\left(\frac{x-x_{FC}}{x_{FC}}\right)^2. \eeq
At $T\ll e_0(x)$ and above the critical point the effective mass
$M^*(x)$ is finite, the energy scale $E_0$ given by Eq. (4) vanishes
and the system exhibits the LFL behavior.
At temperatures $T \geq e_0(x)$ the effective mass $M^*$ starts to
depend on the temperature and the NFL behavior is observed.
Thus, at $|x-x_{FC}|/x_{FC}\ll 1$ the system can be considered as a
highly correlated one: at $T\ll e_0(x)$, the system is LFL, while
at temperatures $T \geq e_0(x)$,  the system possesses the NFL
behavior.  Then, it is clear that at $T\to 0$ the WF law is
preserved.  At $|x-x_{FC}|/x_{FC}\ll 1$, the effective mass given by
Eq. (9) is very large, the Kadowaki-Woods ratio $A/\gamma_0^2$ is
obeyed and the resistivity exhibits the $T^2$ behavior as it was
demonstrated within a simple model of highly correlated liquid
\cite{ksch}.  On the other hand, at $T \geq e_0(x)$, strong
deviations from the $T^2$ behavior occur. We suppose that the
resistivity follows a $T^{\alpha}$ dependence with $1<\alpha<2$ at
$T \geq e_0(x)$.  Here $\alpha=1$ corresponds to strongly correlated
liquid with FC, and $\alpha=2$ corresponds to LFL \cite{ms,ksch}.  We
remark that the outlined behavior was observed in several heavy
fermion metals \cite{vn}.  When the system's density $x$ is outside
the region $|x-x_{FC}|/x_{FC}\ll 1$, the scale $e_0$ becomes
comparable with the Fermi level, the effective mass becomes $M^*\sim
M$ and is approximately constant at energies, $(\mu-\varepsilon)\leq
e_0$. Therefore the system becomes a normal Landau Fermi liquid.

We can expect to observe such a highly correlated
electron (or hole) liquid in heavily overdoped high-T$_c$
compounds which are located beyond the superconducting dome.
Let us recall that beyond the FCQPT point the superconducting gap
$\Delta_1$ can be very small or even absent \cite{ams3}.
Indeed, recent experimental data have shown that this liquid
does exist in heavily overdoped non-superconducting
La$_{1.7}$Sr$_{0.3}$CuO$_4$ \cite{nakam}. Note, that
up to $T=55$ K the resistivity exhibits the $T^2$ behavior,
while at $T\geq 100$ K the resistivity follows a $T^{1.6}$
dependence \cite{nakam}. Thus, we can estimate that
$e_0(x)\sim 50$ K.

Now consider $M^*(B)$ as a function of a weak external
magnetic field $B$ at finite temperatures.
The density $x$ belongs to the
area $(x-x_{FC})/x\ll 1$, and the electron system in
question is the highly correlated one.
The case when the system
has undergone FCQPT was studied in \cite{pogsh}.
This consideration will be applicable to any
2D or 3D electronic Fermi systems. The application of magnetic field
$B$ leads to a weakly polarized state, or Zeeman splitting, when some
levels at the Fermi level are
occupied by spin-up polarized quasiparticles. The width
$\delta p=p_{F1}-p_{F2}$ of the area in
the momentum space occupied by these quasiparticles is of the order
\beq \frac{p_F\delta p}{M^*}\sim B\mu_{eff}.\eeq
Here $\mu_{eff}\sim \mu_B$ is the electron magnetic
effective moment, $p_{F1}$ is the Fermi momentum of the spin-up
electrons, and $p_{F2}$ is the Fermi momentum of the spin-down
electrons. As a result, the Zeeman splitting leads to the change
$\Delta x$ in the density $x$ \beq \frac{\Delta x}{x_{FC}}\sim
\frac{\delta p^2}{p_F^2}. \eeq We assume that $\Delta x/x_{FC}\ll 1$.
Now it follows from Eqs. (9) and (14) that
\beq M^*(B)\sim M\left(\frac{\varepsilon_F}
{B\mu_{eff}}\right)^{2/3}. \eeq
We note that $M^*$ is determined by Eq. (15) as long as
$M^*(B)\leq M^*(x)$, otherwise we have to use Eq. (9).
It follows from Eq. (15) that the application of magnetic field
reduces the effective mass. At finite temperatures $T\leq T^*(x)$,
the effective mass is given by Eq. (15). At temperatures
$T\sim T^*(x)$, both the magnetic field and temperature contribute
to the decreasing of $M^*$. At $T^*(x)\ll T$, the effective mass
is a diminishing function of the temperature. It is clear from Eq. (15)
that $M^*(B)$ remains finite even at $x\to x_{FC}$ and $T\to 0$.
In that case the effective
mass $M^*(x)$ in Eq. (11) has to be substituted by $M^*(B)$.
Note, that if there exists an itinerant magnetic order in the system
which is suppressed by magnetic field $B=B_{c0}$, Eq. (15) has to
be replaced by the equation \cite{pogsh},
$$M^*(B)\sim \left(\frac{1}
{B-B_{c0}}\right)^{2/3}.$$
The coefficient $A(B)\propto (M^*(B))^2$ diverges as
$$A(B)\propto \left(\frac{1}
{B-B_{c0}}\right)^{4/3}.$$

The  behavior described above can be visualized by measuring the
magnetic susceptibility $\chi(T)$ at finite magnetic field so that
$M^*(x)\geq M^*(B)$. The function $\chi(T)$ is a decreasing function of
$T$ at $T\sim e_0$. At $T\ll e_0$, the function $\chi(T)$ becomes
independent of $T$ and starts to depend on $B$, being a decreasing
function of $B$.

Let us comment briefly on the problem of realization of the highly
correlated liquid in dilute Fermi gases and in a low density neutron
matter.  We consider an infinitely extended system composed of Fermi
particles, or atoms, interacting by an artificially constructed
potential with the desirable scattering length $a$.  These objects my
be viewed as trapped Fermi gases, which are systems composed of Fermi
atoms interacting by a potential with almost any desirable scattering
length, similarly to that done for the trapped Bose gases, see e.g.
\cite{nat}.  If $a$ is negative the system becomes unstable at
densities $x \sim |a|^{-3}$, provided the scattering length is the
dominant parameter of the problem. It means that  $|a|$ is much
bigger than the radius of the interaction or any other relevant
parameter of the system. The compressibility  $K(x)$ vanishes at the
density $x_{c1}\sim |a|^{-3}$, making the system be completely
unstable \cite{ams5}.  Expressing the linear response function in
terms of the compressibility \cite{lanl1},
\begin{equation}
\chi (q\rightarrow 0,i\omega \rightarrow 0)
=-\left(\frac{d^{2}E}{d\rho ^{2}}
\right)^{-1},
\end{equation}
we obtain that the linear
response function has a pole at the origin,
$q\simeq 0,\,\omega\simeq 0$, at the same point $x_{c1}$.
To find the behavior of the effective mass $M^*$ as a function of the
density, we substitute Eq. (7) into Eq. (10) taking into account
that $x_{cdw}$ is substituted by $x_{c1}$,
and $q_c/p_F\ll 1$ due to Eq. (16). At low
momenta $q/p_F\sim 1$, the potential $v(q)$ is attractive because
the scattering length is the dominant parameter and negative.
Therefore, the integral on the right hand side of Eq. (10) is
negative and diverges at $x\to x_{1c}$. The above consideration can
be applied to clarification of the fact that effective mass $M^*$ is
again given by Eq. (9) with $x_{FC}<x_{c1}$.  Note that the
superfluid correlations cannot stop the system from squeezing, since
their contribution to the ground state energy is negative. After
all, the superfluid correlations can be considered as additional
degrees of freedom which can therefore only decrease the energy.  We
conclude that the highly correlated behavior can be observed in traps
by measuring the density of states at the Fermi level which becomes
extremely large as $x\to x_{FC}$. At these densities the system
remains stable because of $x_{FC}<x_{c1}$.  It seems quite probable
that the neutron-neutron scattering length ($a\simeq -20$ fm) is
sufficiently large to make it the dominant parameter and  to permit
the neutron matter to have an equilibrium energy, equilibrium
density, and the singular point $x_{c1}$ at which the compressibility
vanishes \cite{ams4}. Therefore, we can expect FCQPT takes place in a
low density neutron matter leading to the stabilization of the matter
by lowering its ground state energy.  A more detailed analysis of
this issue will be published elsewhere.

To conclude, we have shown that our simple model based on FCQPT can
explain the main features of the highly correlated liquid observed in
different Fermi liquids. Thus, FCQPT can be viewed as an universal
cause of the highly correlated behavior.

I am grateful to CTSPS for the hospitality during
my stay in Atlanta. I also thank G. Japaridze for fruitful
discussions.  This work was supported in part by the Russian
Foundation for Basic Research, No 01-02-17189.

\end{document}